# Evidence for Near Ambient Superconductivity in the Lu-N-H System

Nilesh P. Salke[1], Alexander C. Mark[1], Muhtar Ahart[1], Russell J. Hemley[1, 2, 3]

[1]Department of Physics, [2]Department of Chemistry, [3]Department of Earth and Environmental Sciences, University of Illinois Chicago, Chicago, IL 60607 USA

The possibility of superconductivity at ambient pressure and temperature has been a long-standing dream in condensed matter science. The past few years have been witness to significant breakthroughs in this area with the discovery of novel hydrogen-rich materials exhibiting superconductivity in the vicinity of room temperature under pressure.[1-6] In this effort, experiments performed by different groups and broad agreement of the reported very high critical temperatures ($T_c$) have been key to advancing the field. For example, the discovery of the superconducting superhydride LaH$_{10}$ with its $T_c$ of 260 K at megabar pressures by Somayazulu et al.[1] was subsequently confirmed by Drozdov et al.[2] Later, several groups reported $T_c$ at comparable pressures in related Y-H superhydrides,[3-5] and recent work has confirmed the discovery of superconductivity in the C-S-H system at similar temperatures but at lower, but still megabar, pressures.[6]

Recently, Dasenbrock-Gammon et al.[7] reported room-temperature superconductivity of 294 K at a remarkably low pressure of 10 kbar in the Lu-N-H system. However, several groups have reported they are unable to reproduce these results and have asserted that there is no superconductivity in the Lu-N-H system.[8-11] In addition, theoretical studies have not been able to explain these results (*e.g.,* Ref. 12). Measurements of zero electrical resistance and perfect diamagnetism are required to prove superconductivity in a material, and both independently confirmed by other groups. To this end, we have carried out a series of studies of phases in the Lu-N-H system at the conditions of the reported very high $T_c$ superconductivity. As a first step in resolving this controversy, here we report electrical transport measurements on Lu-N-H that closely match both the previously reported $T_c$ values near room temperature and the pressure dependence of the critical temperature.[7]

We obtained samples synthesized by the University of Rochester group and prepared for electrical resistance measurements in diamond anvil cells (DACs) as described by Dasenbrock-Gammon et al.[7] We independently characterized the electrical transport and other properties of the samples in our laboratory at the University of Illinois Chicago and at facilities at Argonne National Laboratory. The four-probe electrical resistance measurements were carried out as a function of temperature at four different pressures from 8.5 kbar to 23 kbar (see Fig. 1). Because of the small size of the sample (and therefore the signal), the measurements were performed using phase-sensitive detection, *i.e.*, with a lock-in amplifier to monitor the magnitude of the voltage across the sample. The signal was then converted to a resistance amplitude $|R|$, which is plotted in Fig. 1. In all runs, sharp drops of more than 5 orders of magnitude were observed at critical temperatures, with a maximum $T_c$ of 276 K at 15 kbar. Technical details on the experimental setup and resistance calculations are provided in the Supplementary Information (SI).

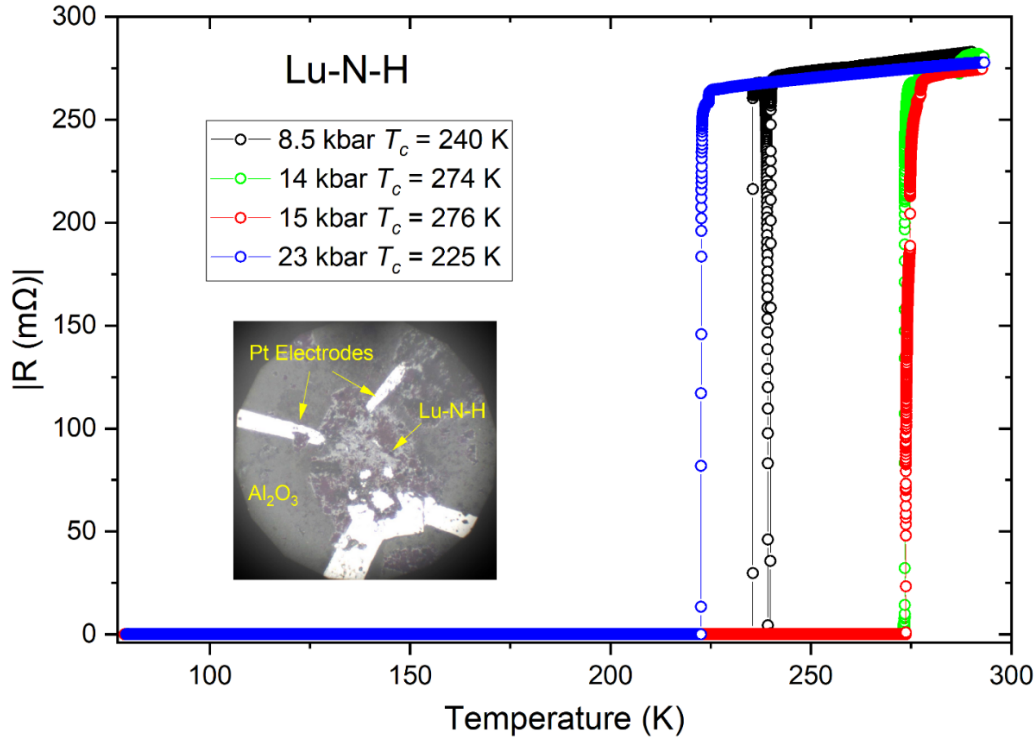

*Figure 1. Temperature dependence of resistance |R| of a Lu-N-H sample at various pressures. Multiple drops in resistance at 8.5 kbar can be attributed to initial sample inhomogeneity that was reduced with subsequent thermal cycling (see SI). The inset shows an image of the sample in a DAC at 2 kbar.*

Figure 2 compares the pressure dependence of $T_c$ of this sample with the data reported by Dasenbrock-Gammon et al.[7] There is good agreement between the two data sets, with small differences attributed pressure gradients that were directly measured from different ruby grains in the sample chamber. *In situ* confocal Raman measurements of the sample in the DAC are consistent with Compound A identified in Ref. 7 as *Fm*-3*m* $LuH_{3-x}N_y$ (see SI). In addition, these spatially resolved spectra collected after transport measurements at 23 kbar showed little variation across the sample, indicating a single dominant phase.

In contrast to the above results, Ming et al.[8] reported measurements on Lu-N-H samples they had synthesized and found no evidence of superconductivity over a temperature range of 10-350 K and pressures up to ~ 40 GPa. The synthesis method described in Ref. 8 used $NH_4Cl$ and $CaH_2$ with a 2:8 ratio as nitrogen and hydrogen sources to react with lutetium metal at 20 kbar and 573-623 K, whereas Dasenbrock-Gammon et al.[7] described using a 1:99 nitrogen/hydrogen gas mixture to react with lutetium foil at 20 kbar and 338 K. On the other hand, Zhang et al.[10] reported using the synthesis method described in Ref. 7, but also found no evidence for superconductivity. Dasenbrock-Gammon et al.[7] reported that during the course of their study of the Lu-N-H system, the synthesis success rate for producing superconducting samples by the described method was 35%. Altogether, these results indicate successful synthesis of the superconducting material is strongly dependent on details of sample preparation, and that further study and optimization of these procedures are needed.

Additional transport measurements provide insight on this sample variability. $R$-$T$ measurements on another sample obtained from the University of Rochester group are shown in Fig. 3. Similar to the results of Ming et al.[8], no abrupt drop in $|R|$ was observed above 80 K. This sample showed $T_c$ of 240 K at 8 kbar and was synthesized in 2021 and had apparently degraded over a period of time. Interestingly, the non-superconducting samples measured in this work and reported by Ming et al.[8] exhibit a resistance maximum as a function of temperature. In our sample, the maximum $|R|$ is observed at 230 K, below which resistance gradually decreases. A similar trend is observed for all measurements in the curves in Ref. 8. This turnover suggests an intriguing interplay of insulator-metal transitions as a function of temperature in the Lu-N-H system reminiscent of those found in early studies of lanthanide hydrides at ambient pressure.[13]

Superconductivity in Lu-N-H could be sensitive to the concentration of nitrogen doping. For hydrides, nitrogen-doping has been shown to be a promising approach to enhance the superconducting properties. Evidence for variable $T_c$ in LaH$_{10}$ may be due to different degrees of nitrogen doping, which can enhance $T_c$ to room temperature.[1, 14] Our recent theoretical study indicates that the electronic properties expected to give rise to the high $T_c$ superconductivity are strongly dependent on the structure, stoichiometry, and N-H-vacancy ordering of the material, *i.e.*, in specific superlattices based on *Fm*-3*m* LuH$_{3-x}$N$_y$.[15] This sensitivity of the electronic structure with respect to crystal structure and atomic ordering indicates that creating the superconductor may be strongly dependent on synthesis conditions, including annealing, consistent with the suggestions above. Additional routes for synthesis of this intriguing material are being explored along with additional characterization and tests of superconductivity.

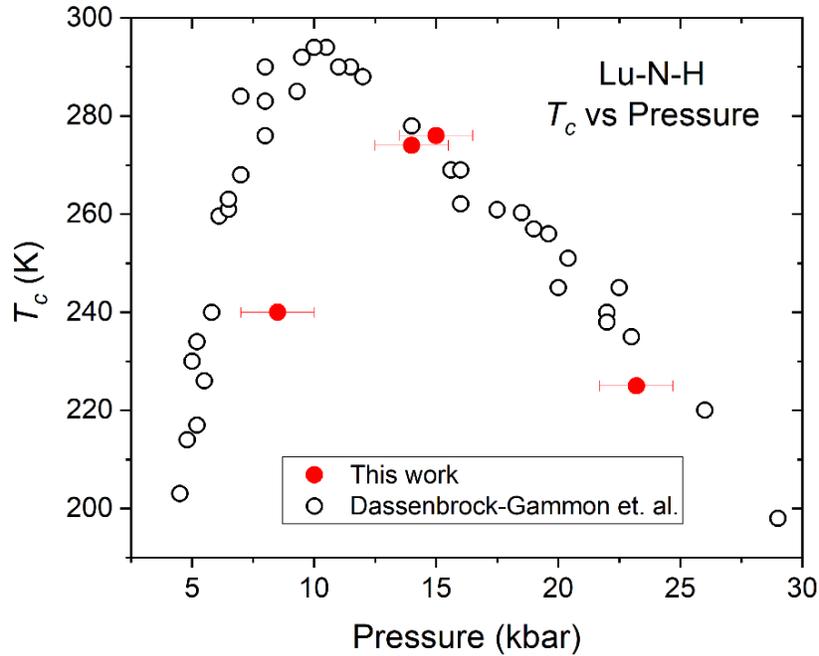

***Figure 2.*** *Pressure dependence of $T_c$ in the Lu-N-H system from this work (red) and reported by Dasenbrock-Gammon et al.[7]*

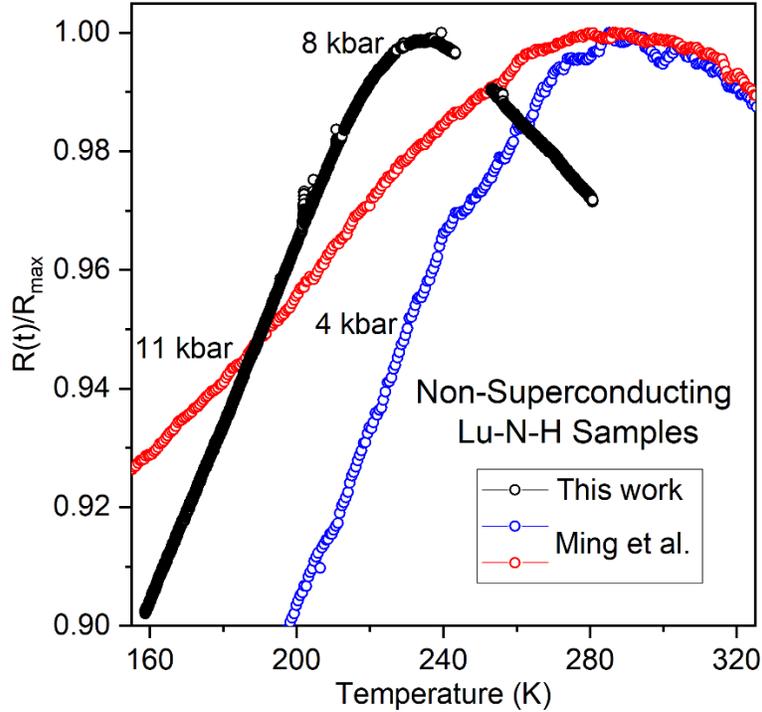

*Figure 3.* Normalized resistance measurements of Lu-N-H samples in which the resistance was not observed to drop to zero. This work (black) 8 kbar data, $R_{max}$ of 414 m$\Omega$ at 240 K and $R_{min}$ of 373 m$\Omega$ at 159 K. The gap in the data is due to the presence of a large amount of environmental noise in the laboratory. Ming et al. (red) 11 kbar, $R_{max}$ of 86.9 m$\Omega$ at 306 K and $R_{min}$ of 78.3 m$\Omega$ at 28.0 K; Ming et al. (blue) 4 kbar data, $R_{max}$ of 237 m$\Omega$ at 274 K and $R_{min}$ of 185 m$\Omega$ at 24.8 K.[8]

**Acknowledgements**

We are grateful to R. Dias for providing the Lu-N-H samples, Q. Li and H. Wen for sharing their data files, D. Kuntzelman and R. Sellers for help with the equipment, and Y. J. Ryu for help with the Raman measurements at GSECARS. This research was supported by the NSF (DMR-2104881) and DOE-NNSA through the Chicago/DOE Alliance Center (DE-NA0003975). Use of the GSECARS Raman Lab System was supported by the NSF (EAR-1531583).

**Data Availability**

Data supporting the findings of this study are available from the public link https://zenodo.org/record/8020419.

# Supplementary Information

**Experimental Details**

We first discuss the Raman spectra of the Lu-N-H sample, which were recorded with a 660 nm excitation laser at GSECARS (Sector 13, Advanced Photon Source, Argonne National Laboratory). Figure S1 shows examples of spectra collected from various locations in the Lu-N-H sample as indicated in the image. The Raman spectra of the Lu-N-H sample match those identified as Compound A shown in the Extended Data Figure 1a in Dasenbrock-Gammon et. al.[1]

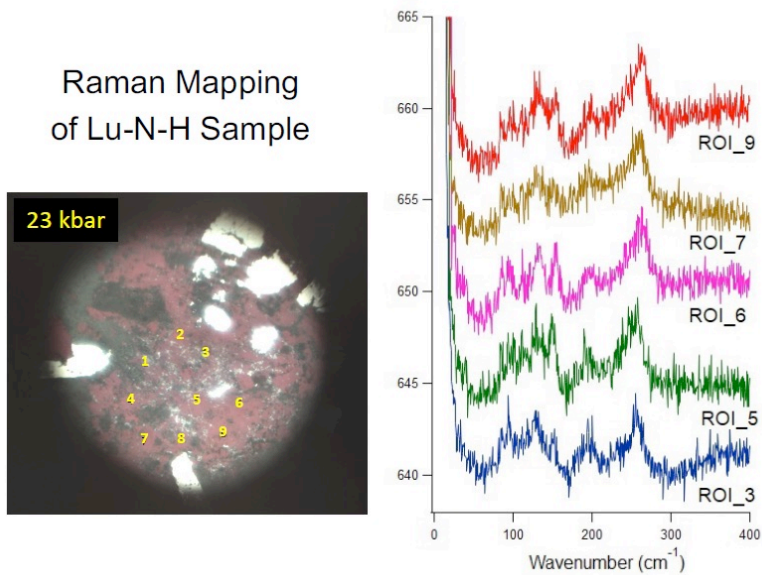

***Figure S1.*** *Raman spectra map of superconducting Lu-N-H sample measured in this study. The reflecting light image shows the Lu-N-H sample at 23 kbar and the Raman measurement locations. Highly reflecting areas correspond to places where the electrodes are in direct contact with the upper diamond anvil.*

In all runs, four electrical contacts were connected to the edges of the sample within the DAC (Fig. 1, inset), with a fifth probe connected to the gasket to check for possible shorts to the cell. The resistance between probes was on the order of 5 Ω, and the resistance from probes to the gasket and to the pressure cell body overloaded the multimeter ($R > 1$ GΩ). The cell was held in a custom vacuum-insulated cryostat (Cryo-Industries) with a copper ring with liquid nitrogen continuously flowing through channels in the sample mount. During cooling the liquid nitrogen flow rate was adjusted to maintain a cooling rate of approximately 0.25 K/minute. During measurements on warming, the liquid nitrogen flow was shut off and the sample was allowed to warm up naturally at approximately 0.3 K/minute.

Thermal insulation was provided using a molecular drag pump (Adixen Drytel 1025) to reduce the pressure inside the cryostat to <0.1 mTorr as measured using a standard gauge (Convectron) and vacuum gauge controller (Granville-Phillips 307). Temperature was measured using two diode sensors, with one mounted on the cell body and one mounted on the copper cooling ring. During cooling, the copper ring cooled much faster than the DAC resulting in a temperature gradient between the ring and sample. Differences of up to 20 K were observed. During warming, the temperature difference between the cell and copper ring did not exceed 2 K, indicating that temperatures recorded during warming were representative of the actual sample temperature. Figure S2 shows the temperature dependence of the resistance during cooling and warming cycles at 8.5 kbar with a noticeable thermal lag during cooling.

The pressure was measured using the standard ruby fluorescence method at ambient temperature after the completion of the warming cycle. The *R-T* relation at 8.5 kbar indicates two large drops in resistance which were not present in higher pressure measurements (Fig. S2). We observed similar behavior in experiments on $LaH_{10}$.[2] We attribute this effect to initial sample inhomogeneity where the local current density in a superconducting region of the sample near the transition exceeds the bulk critical current density of the material, destroying superconductivity. Upon further cooling, the critical current density is expected to increase,[3, 4] thereby further reducing the effect.

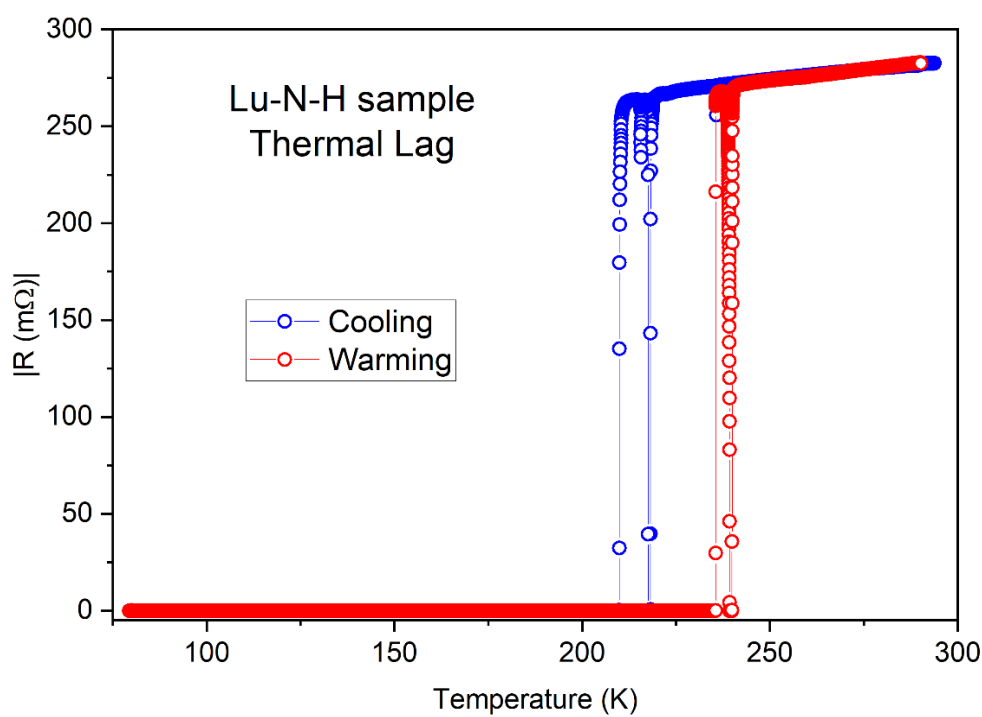

*Figure S2. Comparison between cooling and warming cycles at 8.5 kbar. Only data collected during warming were used in the analysis.*

A current-regulated AC signal was injected into two adjacent electrical contacts. The signals were generated using a DC and AC current source (Keithley 6221) with a sinusoidal waveform and zero DC offset with amplitudes ranging from 7.5 µA to 750 µA and a frequency of 17 Hz. A differential voltage signal was measured across the other two adjacent probes. The measured signal was fed into a high input-impedance transformer preamplifier (Stanford Research SR554) with the gain set to 500X. The preamplifier signal was then sent to a 500 kHz DSP lock-in amplifier (Stanford Research SR860). The sample itself was not electrically connected to earth ground, *i.e.*, for the superconducting sample all electrical measurements were differential and floating with respect to the building ground. The cryostat body and the measuring equipment were referenced to the building ground for shielding.

For the non-superconducting sample shown in Fig. 3, the current was provided by connecting one contact to a voltage-regulated AC output with an amplitude of 500 mV and a frequency ranging from 17 Hz – 1 kHz. An adjacent contact was connected to ground through a 1 kΩ resistor, which was several orders of magnitude larger than the sample resistance, ensuring the current was essentially constant (±0.5%). The transformer preamplifier was not used, which resulted in more environmental noise.

The SR860 lock-in amplifier was fed a reference modulation frequency from the Keithley 6221 internal oscillator. The time constant for the integration stage was set to 1 second and the internal low-pass filter was set to 24 dB. The lock-in amplifier was set to automatically adjust the range and sensitivity of the measurement. For the normal (non-superconducting) state the voltage from the pre-amplifier was on the order of 100 mV for measurements using a 750 µA current, and 10 mV for those using the 75 µA current. The signal in the superconducting state was below the noise floor of the instrument with a measured value on the order of $10^{-6}$ mV. All data was collected using a custom LabVIEW program and no post-processing of the data was performed.

**Comparison with Previous Results**

Figure S3 compares *R-T* curves in this study with previous results by Dasenbrok-Gammon et al.[1] The sample we measured had significantly higher resistance in the normal state compared to those reported in the previous paper.[1] Different ambient temperature resistance values are attributed to different sample sizes. The noise in the resistance oscillates around some finite positive offset arising from the nature of the measurement which utilized a lock-in amplifier setup to record the resistivity as opposed to a pure DC (*e.g.*, van der Pauw) measurement. The lock-in amplifier reports the amplitude of the measured signal |*R*|, resulting in measurements clustering around a positive offset on the order of the absolute value of the noise floor instead of around zero resistance. This is discussed in further detail below.

Because of the low signal levels associated with the small DAC sample and low current required for the measurement, phase-sensitive detection was used to monitor a voltage across the sample with a lock-in amplifier. This method was used, for example, in similar to methods employed in Ref. 1 and some experiments described in Ref. 2. Though lock-in detection greatly suppresses noise, it introduces a phase ambiguity when only the magnitude of the voltage |Δ*V*| difference is recorded. Using |Δ*V*| to compute the resistance (|*R*|) results in a positive offset being observed even in a zero-resistance material (*i.e.,* superconductor). This offset scales with the noise level.

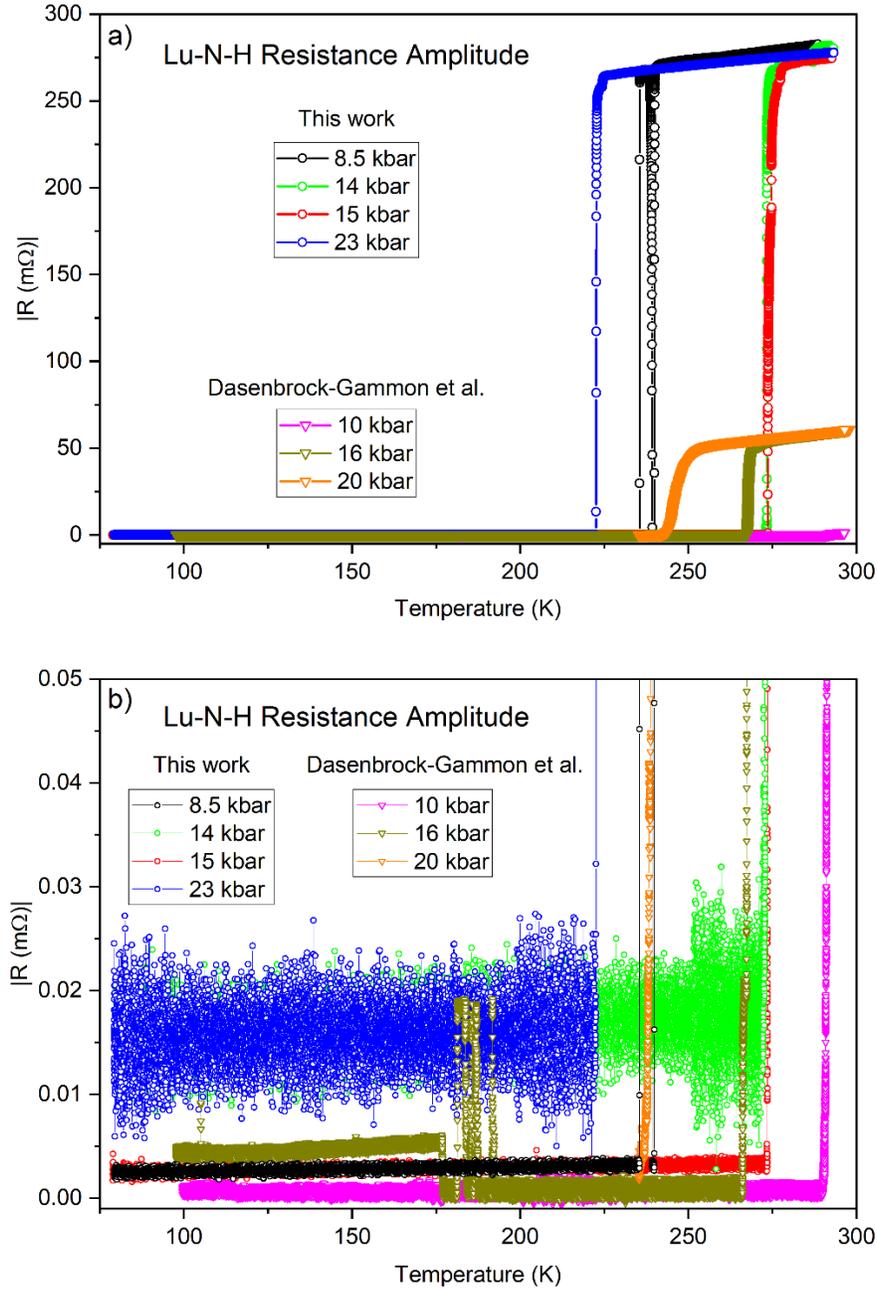

*Figure S3. Temperature dependence of resistance |R| from this work compared with those reported by Dasenbrock-Gammon et. al.[1] a) Full scale. b) Magnified scale near zero |R|.*

Phase-sensitive detection devices such as lock-in amplifiers exploit the fact that sinusoidal signals are orthogonal up to a phase.[5,5] Lock-in amplifiers extract the amplitude ($|A|$) and phase shift ($\theta$) of a signal modulated with a known carrier wave, greatly increasing the signal to noise ratio (SNR) even in extremely noisy environments. In this study a sinusoidal current, $I(t) = |I_{in}| \sin(\omega t)$, was injected into the sample. The magnitude of the AC voltage between two edges of the sample that were in-phase ($V_{ip}$), and in quadrature ($V_{qd}$) where $\theta = 90°$, were measured. Ohm's law was used to calculate in-phase and quadrature resistances. A complex resistance can then be defined, $|R|e^{i\theta} = R_{ip} + iR_q$, where the magnitude $|R|$ is typically reported as the resistance.

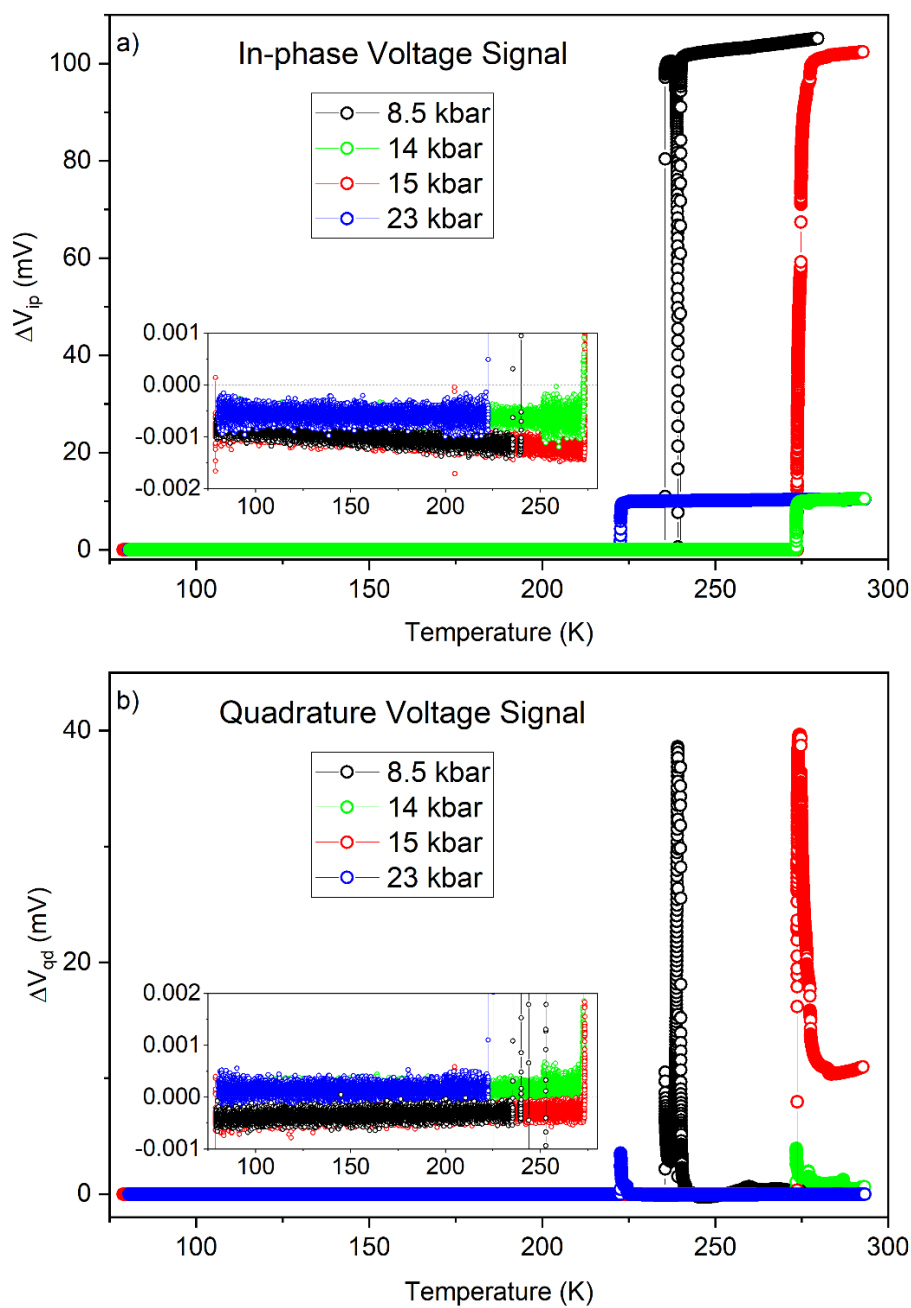

*Figure S4. (a) Temperature dependence of the in-phase component of the complex resistance at various pressures. Inset: superconducting region over a smaller signal range. (b) Temperature dependance of the quadrature component of the complex resistance at various pressures. Inset: superconducting region over a smaller signal range.*

Being a vector magnitude, |R| is positive definite. If the SNR >> 1, as is typically the case with measurements of ohmic materials, this has no impact on resistance measurements. As the SNR approaches 1, that noise can result in random fluctuations causing a negative $\Delta V$. In this situation, using |R| as a measure of the physical resistance is inaccurate; instead of fluctuating around the actual resistance, the data fluctuates around a positive offset. As the SNR drops below 1 this offset increases. In a superconductor the drop in voltage across a sample is 0, which results in a $SNR = \frac{\Delta V}{V_{noise}} \ll 1$ irrespective of how little noise is present in the measurement. Instead of the resistances fluctuating around zero, using $|\Delta V|$ to compute |R| results in values that are always positive and clustered around an offset proportional to the noise level of the measurement (Fig. S3b).

The temperature dependance of the in-phase and quadrature voltages reveal the signals approaching and crossing zero at $T_c$ (Fig. S4). The negative offset in $V_{ip}$ is sensitive to the noise level in the setup. Increasing the magnitude of the carrier current is observed to both reduce the noise and shift the offset closer to zero. The 8.5 kbar and 15 kbar runs (Fig. S3b) were performed using a modulation current amplitude of 750 µA, greatly reducing the noise and causing the in-phase resistance to center around zero. For the 23 kbar and 14 kbar experiments a current of 75 µA was used, resulting in more noise that distorted the signal.